  \documentstyle[12pt]{article}

\catcode`@=11
\def\dif{{\rm d}}
\def\deriv{\@ifnextchar[{\@deriv}{\@deriv[]}}
\def\@deriv[#1]#2#3{\mathchoice%
{{\dif^{#1}#2\over\dif{#3}^{#1}}}{{\dif^{#1}#2/\dif{#3}^{#1}}}%
{{\dif^{#1}#2\over\dif{#3}^{#1}}}{{\dif^{#1}#2/\dif{#3}^{#1}}}}

\def\presup#1{{}^{#1}\kern-.15em\relax}      
\def\presub#1{{}_{#1}\kern-.12em\relax}      

\def\secteqno{\@addtoreset{equation}{section}%
\def\theequation{\thesection.\arabic{equation}}}
\def\endsecteqno{\def\theequation{\@ifundefined{chapter}%
{\arabic{equation}}{\thechapter.\arabic{equation}}}}
\newcounter{subequation}
\def\thesubequation{\alph{subequation}}
\def\sneqnarray{\stepcounter{equation}\let\@currentlabel=\theequation
\setcounter{subequation}{1}
NG\def\@eqnnum{{\rm (\theequation\thesubequation)}}
\global\@eqcnt\z@\tabskip\@centering\let\\=\@eqncr\let\@@eqncr=\@@sneqncr
$$\halign to \displaywidth\bgroup\@eqnsel\hskip\@centering
 $\displaystyle\tabskip\z@{##}$&\global\@eqcnt\@ne
 \hskip 2\arraycolsep \hfil${##}$\hfil
 &\global\@eqcnt\tw@ \hskip 2\arraycolsep $\displaystyle\tabskip\z@{##}$\hfil
  \tabskip\@centering&\llap{##}\tabskip\z@\cr}
\def\endsneqnarray{\@@sneqncr\egroup $$\global\@ignoretrue}
\def\@@sneqncr{\let\@tempa\relax
   \ifcase\@eqcnt \def\@tempa{& & &}\or \def\@tempa{& &}
   \else \def\@tempa{&}\fi
     \@tempa \if@eqnsw\@eqnnum\stepcounter{subequation}\fi
     \global\@eqnswtrue\global\@eqcnt\z@\cr}
\def\nobiblabels{\def\@lbibitem[##1]##2{\@bibitem{##2}}}
\catcode`@=12
\secteqno
\newcommand{\be}{\begin{equation}}
\newcommand{\ee}{\end{equation}}
\newcommand{\bea}{\begin{eqnarray}}
\newcommand{\eea}{\end{eqnarray}}
\newcommand{\bref}[1]{(\ref{#1})}
\newcommand{\ep}{\epsilon}
\newcommand{\vep}{\varepsilon}
\newcommand{\T}{\theta}

\newcommand{\D}{\delta}
\newcommand{\A}{\alpha} \newcommand{\B}{\beta}
\newcommand{\G}{\gamma} 

\def\pa{\partial}

\newcommand{\spc}{\varpi}

\newcommand{\m}{\mu}            
\newcommand{\r}{\rho}           \newcommand{\s}{\sigma}
          
\newcommand{\w}{\omega}
\newcommand{\h}{\eta}
            
\newcommand{\Th}{\Theta}
           
\newcommand{\W}{\Omega}         
\newcommand{\nn}{\nonumber}
\newcommand{\sW}{{\scriptscriptstyle W}}

\def\Diff{diffeomorphism }

\def\CA{{\cal A}}
\def\CM{{\cal M}}

\def\t{\tilde}

\def\WZCC{Wess Zumino consistency condition~}
\font\fiverm=cmr10

\title{{\bf Super-Weyl Invariant  2D Supergravity,
\\ 
Anomaly and WZ Action}}
\author{
        {\sc K.Kamimura$^\dagger$  and}
        {\sc R.Kuriki}$^{\dagger\dagger}$\\
\\
        \llap{$^\dagger$}%
        \small{\fiverm{Department of Physics, Toho University}}\\
        \small{\fiverm{Funabashi,}}
        \small{\fiverm{274 JAPAN}}\\
        \llap{$^{\dagger\dagger}$}%
        \small{\fiverm{Ochanomizu University,
        Department of Physics, 
        Faculty of Science}}\\
        \small{\fiverm{Otsuka-Machi, 
         Bunkyo-Ku}}
        \small{\fiverm{Tokyo,}}
        \small{\fiverm{112 JAPAN}}\\
 \small{
kamimura@ph.sci.toho-u.ac.jp,
r-kuriki@fs.cc.ocha.ac.jp}\\
}
\date{}

\begin{document}

\maketitle
\vskip 5mm

\thispagestyle{empty}
\begin{abstract}
We present a candidate of anomaly and Wess Zumino action of the 
two dimensional supergravity coupling with matters
in a super-Weyl invariant regularization. 
It is a generalization of the Weyl and the area preserving \Diff 
invariant formulation of
two dimensional gravity theory.
\end{abstract}

\vfill
\vbox{
\hfill July 1997 \null\par
\hfill TOHO-FP-9755 \null\par
\hfill OCHA-PP-102  \null\par
}\null
\clearpage


\section{Introduction}
\indent

Recently two dimensional gravity theory coupling with scalar matter 
is discussed in the Weyl invariant regularization 
\cite{kmk}\cite{j}\cite{abs}\cite{new1}\cite{new2}\cite{gkk}. 
In the standard diffeomorphism invariant formulation the Weyl 
transformations become anomalous and the gauge parameter
of Weyl symmetry is propagating at the quantum level
\cite{p1}. 
In contrast to the regularization which respects
the two dimensional diffeomorphisms,
the Weyl invariant formulation
causes a breaking of the two dimensional diffeomorphisms.
However the two dimensional diffeomorphisms (Diff$_2$) 
is not fully broken but the area preserving \Diff (SDiff) still
remains invariant.
The form of anomaly which possesses
the Weyl and the area-preserving diffeomorphism 
invariances has been constructed.
Moreover a local form of the Wess-Zumino (WZ)
term is obtained using the general formalism 
in terms of the coset coordinate of $\frac{Diff_2}{SDiff}~$\cite{gkk}.
\medskip

In this paper we will extend the discussions
on the Weyl invariant gravity theory to two dimensional (2D)
supergravity theory.
We present a candidate of anomaly satisfying the \WZCC
assuming  super-Weyl invariant reguralizations. 
We also obtain the local WZ action for the anomaly.
Although the \Diff symmetries become anomalous 
the SDiff remains unbroken as in the bosonic case. 
In \cite{ka} the author investigated an extension of area-preserving 
structure to 2D supergravity and conjectured a 
non-local form of effective action invariant under the super 
extension of area-preserving \Diff.
However the super-Weyl invariant anomaly which we will give in this 
paper does not allow such super extension of the SDiff.
The two dimensional local supersymmetries are fully anomalous in 
the present super Weyl invariant formulation. 
\medskip

This paper is organized as follows.
In section 2 we give a brief review 
of the 2D supergravity 
invariant under \Diff and the two dimensional supersymmetry.
The super-Weyl
anomaly and WZ action are also given there.
In section 3 
we introduce some super-Weyl invariant variables
using a canonical transformation.
The super-Weyl invariant anomaly and WZ action is given in section 4.
We present some discussions in the last section
\vskip 6mm

\section{Super-Weyl anomaly and WZ term}
\indent

Before going into the Super-Weyl invariant formulation of the 
two dimensional supergravity
we briefly review the \Diff invariant formulation since our main discussions
are based on them. We follow the discussion of the reference \cite{gkk1} and 
use the notations.

The two dimensional supergravity theory 
can be formulated by
the Neveu-Schwarz-Ramond superstring  model. 
The classical action is given by
\be
S_0~=~-\int d^2x~e~\biggl[~\frac{1}{2}
\bigl(g^{\A\B}\pa_\A X\pa_\B X-i\overline\psi\r^\A\nabla_\A\psi\bigr)
+\overline\chi_\A\r^\B\r^\A\psi\pa_\B X
+\frac{1}{4}(\overline\psi\psi)(\overline\chi_\A\r^\B\r^\A
\chi_\B)~\biggr],
\label{S0}
\ee
where $X^\m,~ \psi^\m,~e_\A{}^a$ and $\chi_\A$ 
are respectively the bosonic and fermionic string coordinates,
the zwei-bein and gravitino fields. 
The space-time indices is $\mu=0,\ldots,D-1$, which is abbreviated in 
\bref{S0},
and the world-sheet indices is $\alpha=0,1$.
The covariant derivative is defined by
\be
\nabla_\alpha\psi\equiv(\partial_\alpha+\frac{1}{2}
\spc_\alpha\rho_5)\psi,~~~~~~~~
\spc_\A~=~
\frac{1}{e}e_\A^{~a}\ep^{\B \G}\pa_\B e_{\G a}~+~
2 i \bar{\chi}_\A \r_5\r^\B \chi_\B.
\ee
The action has local symmetries under the two dimensional diffeomorphisms, 
the local Lorentz, the two dimensional
world-sheet local supersymmetries, 
the Weyl rescaling, the local fermionic transformations.
The last one is a result from the property of the two dimensional
gamma matrixes.
We will refer the Weyl rescaling and the local fermionic transformations
as `super-Weyl transformations (SW) '.

In the regularization which respects Diff$_2$
and the two dimensional world-sheet local supersymmetries
the SW symmetries cannot be maintained but become anomalous.
The SW anomaly satisfying the WZ consistency condition
$~ \delta{\cal A}~=~0~$, 
has been given by
\bea
{\cal A}&=&-i\hbar k \int d^2x[~R~C_\sW~-~
4~{\bar T}~\r_5~\h_\sW~],
\label{nianomaly}\\
&&R~\equiv~2~(\ep^{\A\B}\pa_\A \spc_\B),~~~~~~~
T~\equiv~\ep^{\A\B}~\nabla_\A~\chi_\B.
\eea
where $~k~$ is a constant 
and  $C_\sW$ and $\h_\sW$ are 
ghost variables  corresponding to the Weyl and the local fermionic
transformations respectively.

It is well-known that we can construct a local counter term  
for the anomaly by a introducing extra field, the super Liouville mode.
The following general method \cite{gps} 
the WZ action is constructed \cite{gkk1} using Wely 
degrees of freedom $\s$ and super-Wely degrees of freedom $\h$ as,
\bea
{\cal M}_1~=~\hbar k~ S^{SL}~=~\hbar k~(~ S^{SL}_0~+~S^{SL}_1),
\label{sla}
\eea
with
\bea
S^{SL}_0&=&-\int d^2x ~e~\Bigl\{\frac{1}{2}
\bigl(g^{\A\B}\pa_\A \sigma\pa_\B \sigma
-i\overline\h\r^\A\nabla_\A\h\bigr)
-\overline\chi_\A\r^\B\r^\A\h\pa_\B \sigma
+\frac{1}{4}\overline\h\h\overline\chi_\A\r^\B\r^\A
\chi_\B \Bigr\},
\\
S^{SL}_1&=&-~\int d^2x
[R~\sigma~-~4~{\overline T}\r_5 \h].
\label{niwzterm}
\eea
The WZ action verifies $ \D \CM_1~=~i \CA~$  {\it off shell} and 
the BRST transformations of the extra variables are given by
\bea
&&\delta_0\sigma~=~\partial_\A
\sigma~{C}^\A-\overline\h\omega-{C}_\sW~,\nn\\
&&\delta_0\h  ~=~\partial_\A\h {C}^\A+\frac{1}{2}\r_5\h {C}_L
+i\r^\A\omega(\partial_\A\sigma+{\overline\h}\chi_\A)-\frac14 \h {C}_\sW-
\eta_{\sW},
\label{deltheta}
\eea
where the ghosts for
the diffeomorphisms, the local Lorentz and the two dimensional
world-sheet local supersymmetries are denoted as
$C_\alpha$, $C_L$ and $\omega$ respectively.
\medskip

The field-antifield formalism is one of the most useful formalism 
to investigate the gauge structure.
The classical master equation
(CME) , $(S,S)=0$,
which is defined in the local functional space of fields and 
antifields, encodes all gauge structure of the theory
\cite{HT}. The minimal proper solution of the CME is
\bea
S~&&=~S_0~
\nn\\&&
+~X^*~(\partial_\A X~C^\A+\overline\psi\w~) \nn \\
&& +~\psi^*~(\partial_\A\psi~C^\A~+~\frac{1}{2}\r_5\psi~C_L
-\frac{1}{4}\psi~C_\sW-i\r^\A\w(\partial_\A X-\overline\psi~\chi_\A))
\nn \\
&& +~{e^\A_{~a}}^*~(\partial_\B e_\A{}^a~C^\B~+~e_\B{}^a\partial_\A~C^\B
+\ep^a{}_b~e_\A{}^b~C_L+\frac{1}{2}~e_\A{}^a~C_\sW
+2i\overline{\chi_\A}~\r^a~\w~)
\nn \\
&& +~{\chi^\A}^*~(~\partial_\B\chi_\A~C^\B
+\chi_\B~\partial_\A C^\B
+\frac{1}{2}\r_5\chi_\A~C_L~+~\frac{1}{4}\chi_\A~C_\sW
+\frac{i}{4}\r_\A~\h_{\scriptscriptstyle W}+
\nabla_\A\w~) \nn \\
&& +~{C_\A}^*~(~\partial_\B C^\A~C^\B
-i\overline\w\r^\A\w ~) \nn \\
&& +~ {C_L}^*~(~\pa_\A C_L~C^\A
+\frac{1}{2}\overline\w\r_5\h_{\scriptscriptstyle W}
-  \spc_\A~i\overline\w\r^\A\w) \nn \\
&& +~{\w}^*~(~\partial_\A\w~C^\A~+~\frac{1}{2}\r_5\w~C_L~+~
\frac{1}{4}\w~ C_\sW ~+~\chi_\A~(i\overline\w\r^\A\w)) \nn \\
&& +~{C_\sW}^*~(~\partial_\A C_\sW~C^\A
-\overline\w\h_{\scriptscriptstyle W} ~) \nn \\
&&+~{\h_{\scriptscriptstyle W}}^*~(
\partial_\A \h_{\scriptscriptstyle W}~C^\A
+\frac{1}{2}\r_5\h_{\scriptscriptstyle W}~C_L
-\frac{1}{4}\h_{\scriptscriptstyle W}~C_\sW
-i\r^\A\w(\partial_\A C_\sW~+~\overline\chi_\A
\h_{\scriptscriptstyle W})
-\frac{4}{e} \w ~(\overline\w \r_5 T)~) \nn\\
&&-~\frac{1}{2 e}(\psi^*\w)(\psi^*\w).
\label{solCME}
\eea
It is easy to read the BRST transformations of the fields. 
It contains a quadratic term of the anti-field $\psi^*$  reflecting 
the fact that the two dimensional local supersymmetry algebra 
on the spinor $\psi$ closes only on shell of its equation of motion.
\smallskip


\section{Super-Weyl invariant variables}
\indent

Here we will consider the SW invariant formulation. 
In order to maintain the manifest SW invariance it is useful to
introduce
the SW invariant variables \cite{ka}
\bea
\t X &=&~X ,~~~~~~~~~~~~~~~~~~
\t\psi~=~{e^{1/4}}~\psi,
\nn\\
\t e_\A^a&=&\frac{1}{e^{1/2}}e_\A^a,~~~~~~~~~~~~
\t\chi_\A~=~\frac{-1}{2 e^{1/4}}\r^\B\r_\A\chi_\B.
\label{invvari}\eea
They are not independent but satisfy identically
\be
{\rm det}~\tilde e_\A^a~=~1,~~~~~~~~~\t\r^\A~\t\chi_\A~=~0,
\ee
where Dirac matrix $\t\r^\A(x)$ is defined using $\t e_\A^a$ and
the indices of $\t\chi^\A$ and $\t\r^\A$'s
are lifted and lowered by SW invariant metric 
$\t g^{\A\B}$ and $\t g_{\A\B}$
defined by $\tilde{e}_\alpha{}^b$ and $\tilde e^\beta{}_a$. 

\vskip 6mm

Since the classical action $S_0$ in \bref{S0} is invariant under
SW transformations it is expressed in terms of the invariant variables
\bea
\t S_0&=&-\int d^2x\biggl[~\frac{1}{2}
\bigl(\t g^{\A\B}\pa_\A \t X\pa_\B \t X-i\overline{\t\psi}
\t\r^\A\t\nabla_\A\t\psi\bigr)-
2~\overline{\t\chi}^\B~\t\psi\pa_\B\t X
-\frac{1}{2}(\overline{\t\psi}\t\psi)(\overline{\t\chi_\B}
\t\chi_\B)\biggr],
\label{tS0}
\eea
where the covariant derivative, which is invariant under SW transformations, 
is
\bea
\t\nabla_\A\t\psi~=~(\pa_\A+\frac12\r_5\t\spc_\A)\t\psi,~~~~~~~~~~~ 
\t\spc_\A~=~\t e_\A^a~\ep^{\r\s}\pa_\r \t e_{\s a}.
\eea

The BRST transformations of new variables 
are obtained from \bref{solCME}
by making a canonical transformation 
in the sense of anti-bracket formalism.
The transformation \bref{invvari} is generated by the generating
function 
\bea
W(\phi,\t\phi^*)&=&\t X^*~X~+~\t\psi^*~(e^{\frac14}~\psi)~+~\t e^{*\A}_a
(\frac{e^a_\A}
{e^{\frac12}})~+~E^*~e~
\nn\\
&+&\t\chi^{*\A}~(~\frac{-1}{2 e^{\frac14}}\r^\B\r_\A\chi_\B~)~+~
\hat\chi^*~(~i~e^{\frac14}~\r^\A~\chi_\A~)
\nn\\
&+&\t C^*_L~(~C_L~-~i~\bar\w~\r_5~\r^\A~\chi_\A~)~+~
\t C^*_\sW~(~C_\sW~+~2i~\bar\w~\r^\A~\chi_\A~)
\nn\\
&+&\t C^*_\A~C^\A~+~\t\w^*~(~\frac{\w}{e^{\frac14}}~)~
+~\t\h^*_\sW~(~{e^{\frac14}}~\h_\sW~).
\eea
It also defines SW invariant ghosts $\t C^\A,~\t\w$ and $\t C_L$ and
SW {\it non-invariant} fields $~E,~\hat \chi,~ \t C_\sW$ and $\t\h_\sW$.
In terms of new variables the solution $S$ of CME in \bref{solCME} becomes
a sum of two terms
\bea
\t S&=&\t S_{inv}~+~\t S_{non}.
\label{tsolCME}
\eea
The first term $\t S_{inv}$ contains only SW invariant variables as
\bea
\t S_{inv}&=&\t S_0~+~
\nn\\&&
+~\t X^*~(\partial_\A \t X~C^\A+\overline{\t\psi}~\t\w~) \nn \\
&& +~\t\psi^*~(\partial_\A\t\psi~C^\A~+~\frac{1}{2}\r_5~\t\psi~\t C_L~
+~\frac{1}{4}~\t\psi~(\partial_\B~C^\B)~-i\t\r^\A\t\w(\partial_\A \t X-
\overline{\t\psi}~\t\chi_\A))
\nn \\
&& +~{\t e}^{\A *}_{~a}~(~\partial_\B \t e_\A{}^a~C^\B
+\t e_\B{}^a\partial_\A~C^\B~-
\frac12 \t e_\A{}^a(\partial_\B~C^\B)
+\ep^a{}_b~\t e_\A{}^b~\t C_L+2i\overline{\t\chi_\A}~\r^a~\t\w~)
\nn \\
&& +~{{\t\chi}^{\A *}}~(~\partial_\B\t\chi_\A~C^\B
+\t\chi_\B~\partial_\A C^\B-\frac{1}{4}\t\chi_\A~(\partial_\B~C^\B)
+\frac{1}{2}\r_5\t\chi_\A~\t C_L
+\t\nabla_\A\t\w-\frac{i}{2} \t\r_\A\W~)
\nn \\
&& +~{C_\A}^*~(~\partial_\B C^\A~C^\B
-i\overline{\t\w}\t\r^\A\t\w ~) \nn \\
&& +~ {\t C_L}^*~(~\pa_\A \t C_L~C^\A
-~\overline{\t\w}~\r_5~\W~-~\t\spc_\A~i\overline{\t\w}\t\r^\A\t\w) \nn \\
&& +~{\t\w}^*~(~\partial_\A\t\w~C^\A~+~\frac{1}{2}\r_5~\t\w~\t C_L~-~
\frac{1}{4}~\t\w~ (\partial_\B~C^\B) ~+~
\t\chi_\A~(i\overline{\t\w}\t\r^\A\t\w)) 
\nn \\
&&-~\frac{1}{2}(\t\psi^*~\t\w)(\t\psi^*~\t\w).
\label{solCMEswi}
\eea
where
\bea
\W~\equiv~i\t\r^\A\t\nabla_\A~\t\w~+~
2~(\overline{\t\chi^\G}\t\chi_\G)~\t\w.
\eea
The remaining terms in $\t S_{non}$ contain the SW non invariant variables.
Since the transformation is canonical, preserving the anti-brackets,
$\t S$ is still satisfying the CME, $(\t S,\t S)=0$. Furthermore 
the first term $\t S_{inv}$ satisfies 
CME $(\t S_{inv},~\t S_{inv})=0~$ by itself.
In other words we have introduced new ghosts $\t C_L,~\t C_\sW, etc.~$ 
so that their BRST transformations are closed. Actually 
if we start from the action $\t S_0$ in \bref{tS0}
we obtain the $\t S_{inv}$ as the solution of CME.

\medskip

\section{SW invariant Anomaly and WZ action}
\indent

The SW invariant anomaly is described by the SW invariant variables.
In order to find it, we first recognize that 
the forms of the BRST transformations of new variables
have the same structures as the corresponding original variables.
$\t S_{inv}$ in \bref{solCMEswi} is obtained from
$ S$ in \bref{solCME}
by the following replacements :
\be
e_\A^a~\rightarrow~\t e_\A^a,~~~~~~~  
\chi_\A~\rightarrow~\t \chi_\A,~~~~~~~  
\w~\rightarrow~\t \w,~~~~~~~  
C_L~\rightarrow~\t C_L,
\label{rep1}
\ee
 and
\be
C_\sW ~\rightarrow~~-~(\pa_\A C^\A),~~~~~~~~~
\h_\sW ~\rightarrow~~-~2~\W.
\label{rep2}\ee
That is the Weyl ghost $C_\sW$ is replaced 
by a combination of diffeomorphism
ghost $-(\pa C)$
and the super Weyl ghost $\h_\sW$ 
is replaced by that of supersymmetry ghost $-2\W$.
Furthermore the transformation properties of $(\pa_\A C^\A)$ and $\W$
are 
\bea
\D(\pa C)&=&\pa_\B(\pa C)C^\B-2\overline{\t\w}\W
\\
\D\W&=&\pa_\A\W C^\A+\frac12 \r_5\W \t C_L+\frac14 \W(\pa C)
-\frac{i}{2}\t\r^\A\t\w(\pa_\A(\pa C)+2\overline{\t\chi_\A}\W)+
2\t\w(\overline{\t T}\r_5\t\w)
\eea
which are also obtained from 
$\D C_\sW$ and $\D\h_\sW$ by the same replacements.

Guided by the transformation properties of the new variables  
one can deduce a candidate of the SW invariant anomaly satisfying 
WZ consistency condition.
Knowing the SW non-invariant anomaly \bref{nianomaly} and 
making the replacements \bref{rep1} and \bref{rep2} we obtain
\be
\CA^{SW}~=-i\hbar k\int{d^2x}(~-
\t R~(\pa C)~+~8~\t T\r_5~\W~),
\label{swia}
\ee
where
\be
\t R~=~2~\ep^{\A\B}\pa_\A\t\spc_\B,~~~~~~~~
\t T~=~\ep^{\A\B}\t\nabla_\A\t\chi_\B.~~~~~~~~
\ee
It is straightforward to show 
that it satisfies the WZ consistency condition
\be
\D~\CA^{SW}~=~0. 
\ee
That is $~\CA^{SW}~$ is the candidate of a SW invariant anomaly 
which would be obtained in the SW invariant regularizations at  
one loop level.

\vskip 6mm

Next we find the WZ action $\CM^{SW}_1$ for the SW invariant 
anomaly \bref{swia}.
It is the solution of 
\be
\D \CM^{SW}_1(\phi,\T)~=~i~\CA^{SW}
\ee
and is found as a function of the classical fields $\phi$ and
the extra variables $\T$. The latter are essentially anomalous 
gauge degrees of freedoms  whose  dynamical action
is the WZ term.

In the case of gravity they are finite \Diff functions $f^\mu(x)$.
Since the SDiff is non anomalous the WZ action is a function of
only one degrees of freedom 
$\Theta$ parametrizing the coset $\frac{Diff}{SDiff}$.
In the present case the local supersymmetry is also anomalous.
The WZ action is expected to  be a function of 
a bosonic extra field $\Theta$ and a fermionic extra spinor $\Psi$. 

The explicit form of the WZ action is obtained from the form of    
the SW non invariant WZ term, i.e. super Liouville
action \bref{sla}-\bref{niwzterm}, 
by the replacements \bref{rep1} and 
\be
\s ~~\rightarrow~~-~\Theta,~~~~~~~~~
\h~~\rightarrow~~-~\Psi.
\label{rep3}
\ee
as
\bea
{\cal M}^{SW}_1~=~\hbar k~ S^{SW}~=~\hbar k~(~ S^{SW}_0~+~S^{SW}_1),
\label{swa}
\eea
with
\bea
S^{SW}_0&=&-\int d^2x[\frac{\t g^{\A\B}}{2}\pa_\A \Th\pa_\B \Th
-\frac{i}{2}\overline\Psi\t\r^\A\pa_\A\Psi+2
(\overline{\t\chi^\B}\Psi)\pa_\B\Th
-\frac12(\overline\Psi\Psi)(\overline{\t\chi^\A}\t\chi_\A)]
\label{wzterm0}\\
S^{SW}_1&=&-\int d^2x[-\t R\Th~+~4\overline{\t T}\r_5 \Psi].
\label{wzterm1}
\eea
It is noted that
$S^{SW}_0$ is obtained from the classical action $\t S_0$ in \bref{tS0}
by the replacements 
$\t X \rightarrow  \Theta$ and $\t\psi \rightarrow - \Psi$
and $S^{SW}_1$ is obtained from the anomaly by the replacements
$(\pa C) \rightarrow \Theta$ and $2\W \rightarrow \Psi$.
The WZ action verifies 
\be
\D~\CM^{SW}_1~=~i~\CA^{SW}~
\ee
{\it off shell}
if the BRST transformations of the extra variables $\Th$ and $\Psi$ are 
introduced as
\bea
\D~\Th&=&\pa_\A\Th C^\A~-~\overline\Psi\w~-~(\pa C),
\\
\D~\Psi&=&\pa_\A\Psi C^\A~+~\frac14\Psi(\pa C)~+~
\frac12\r_5 
\Psi\t C_L~+~i\t\r^\A\w(\pa_\A\Th+\overline\Psi\t\chi_\A)~-~2\W.
\label{trnsextra}
\eea
They are deduced from the SW non invariant ones \bref{deltheta}
by the replacements \bref{rep1}, \bref{rep2} and  \bref{rep3}.

\section{Discussions}
\indent

In this paper we have discussed two dimensional supergravity 
under a SW invariant regularization. We have found
the candidate anomaly and the Wess Zumino action.
They are deduced from the properties of the BRST transformations 
of the SW invariant supergravity multiplet.
The crucial property is the correspondence of the BRST transformations
between the original and the SW invariant variables. The latter is obtained 
from the former by the replacements \bref{rep1} and
\bref{rep2},
\be
C_\sW ~\rightarrow~~-~(\pa_\A C^\A),~~~~~~~~~
\h_\sW ~\rightarrow~~-~2~\W~=~-2~[~i\t\r^\A\t\nabla_\A~
+~2~(\overline{\t\chi^\G}\t\chi_\G)~]~\t\w~.
\label{rep21}\ee
In the anomaly \bref{swia} the \Diff ghosts $C^\A$ 
appears in the combination of $(\pa C)$.
It means that the \Diff transformations with the infinitesimal parameter
$\ep^\A$ satisfying  
\be
(\pa_\A \ep^\A)~=~0, ~~~~~~i.e.~~~~~~~\ep^\A~=~\ep^{\A\B}~
\pa_\B~\ep,
\label{apdiff}
\ee
are not anomalous. The area preserving diffeomorphisms remain
non anomalous as in the bosonic theory. 
On the other hand supersymmetry ghosts $\w$ appears in the combination 
of $\W$. The local supersymmetry transformations would be non anomalous if
the spinor parameter $\t\vep$ is satisfying
\be
[~i\t\r^\A\t\nabla_\A~+2~(\overline{\t\chi^\G}\t\chi_\G)~]~\t\vep~=~0.
\label{apsusy}
\ee
However there found no local solutions of \bref{apsusy} 
unlike the case of \bref{apdiff}. 
Furthermore the WZ action \bref{swa} has no additional gauge symmetry.
In the last paper\cite{gkk} we have shown that there appears additional
gauge invariance of the WZ action if we would have introduced
extra variables associated to the non anomalous transformations.
It follows the degree of freedom of $\Psi$ 
agrees with that of the fermionic anomalous transformations. 
We can conclude that the all 
two dimensional world-sheet supersymmetries become anomalous
under the SW invariant regularization. 
It is also consistent with the fact that 
the number of independent anomalies should be unchanged 
under different regularizations.
It is equal in the SW non-invariant and SW invariant regularizations.

The anomaly \bref{swia} would be obtained 
by the Pauli-Villars regularization
using the counter term
\bea
\t S^M&=&-\int d^2x~\frac{1}{2}~[~M^2~\t X~\t X~-~i~M~
\overline{\t\psi}~\t\psi~].
\label{counter}
\eea
It naturally breaks 
the local supersymmetries and area non-preserving \Diff
while it is invariant under area preserving \Diff, local Lorentz
and SW transformations. 
\medskip

The BRST transformation of the extra variables 
\bref{trnsextra} is nilpotent only on shell of equation of WZ action,
$[S^{SW}]_\Psi=0$ ,
\be
\D^2~\Th~=~0,~~~~~~~~~\D^2~\Psi~=~\t\w(\overline{\t\w}[S^{SW}]_\Psi)~
\sim~0.
\ee
The similar situation happened in the case of SW non-invariant formalism
\cite{gkk1}.
As in the latter case it requires an additional term 
\be
-~\frac{1}{2}(\Psi^*~\t\w)(\Psi^*~\t\w)
\label{afPsi}
\ee
in the quantum action. 
However the reasons of  off shell non nilpotency are quite different. 
In the latter the off shell non nilpotency arises from the fact that 
the non-anomalous transformations ( local susy ) do not close
but give anomalous ( super-Weyl ) transformations.
In the present case the non anomalous transformations close by themselves.
The  off shell non nilpotency comes from the fact that the local supersymmetry
transformations
for the spinor $\Psi$ close only on shell of its equation of motions.
\medskip

It is interesting to find the
relation of the extra variables $\Theta$ and $\Psi$ with
the local gauge degrees of freedoms. In the case of two dimensional
gravity theory
$\Theta$ is expressed by the finite \Diff function $f^\A(x)$ as
\bea
\Th(x)~=~\ln(\Delta^f)_{x=F(x)},~~~~~~{\rm with} ~~~~~~F(f(x))~=~x.
\eea
The general framework to obtain such relations has been established
\cite{gkk}. 
It is also shown that it is the coordinate parametrizing the coset
$\frac{Diff_2}{SDiff}$.
In the present supergravity case the 
extra variables $\Theta$ and $\Psi$ are regarded as the coordinates
parametrizing the coset $\frac{Diff_2\times SUSY_{local}}{SDiff}$. 
The analysis in the general formalism and the explicit form of 
the extra variables in terms of finite gauge transformation
parameters will be discussed elsewhere.

\bigskip

{\bf Acknowledgments}
\medskip

The authors would like to thank 
Prof.J.Gomis and 
Prof.D.R.Karakhanyan for
helpful discussions.

\vskip 6mm


\end{document}